\documentclass[ALICE,manyauthors]{cernphprep}

\usepackage[comma,square,numbers,sort&compress]{natbib}
\usepackage{hyperref}
\usepackage{lineno}
\usepackage{color}
\usepackage{amsmath}

\newcommand{\energy}{\ensuremath{{\sqrt{s_{\mathrm{NN}}}}}} 

\begin{document}%
\setlength{\extrarowheight}{.25em}

\begin{titlepage}
\PHyear{2018}
\PHnumber{095}      
\PHdate{9 May}  
%


\title{Measurements of low-$p_{\rm T}$ electrons from semileptonic heavy-flavour hadron decays
       at mid-rapidity in pp and Pb--Pb collisions\\ at
       $\mathbf{\sqrt{{\it s}_\mathrm{NN}}}$ = 2.76 TeV}
\ShortTitle{$R_{\rm AA}$ of electrons from semileptonic heavy-flavour
            hadron decays} 
\Collaboration{ALICE Collaboration\thanks{See Appendix~\ref{app:collab} for the list of collaboration members}}
\ShortAuthor{ALICE Collaboration} 

\begin{abstract}
 Transverse-momentum ($p_{\rm T}$) differential yields of electrons from semileptonic 
  heavy-flavour hadron decays have been measured in the most central (0--10\%) and in 
  semi-central (20--40\%) Pb--Pb collisions at \energy~ = 2.76 TeV. The corresponding production 
  cross section in pp collisions has been measured at the same energy with substantially 
  reduced systematic uncertainties with respect to previously published results. 
  The modification of the yield in Pb--Pb collisions with respect to 
  the expectation from an incoherent superposition of nucleon-nucleon collisions is 
  quantified at mid-rapidity ($|y|$ $<$ 0.8) in the $p_{\rm T}$ interval 0.5--3 GeV/$c$ 
  via the nuclear modification factor, $R_{\rm AA}$. This paper extends the $p_{\rm T}$ 
  reach of the $R_{\rm AA}$ measurement towards significantly lower values with respect 
  to a previous publication.  In Pb--Pb collisions the $p_{\rm T}$-differential measurements of yields at low $p_{\rm T}$ are essential to investigate the scaling of heavy-flavour production with the number of binary nucleon-nucleon collisions.
Heavy-quark hadronization, a collective expansion and even initial-state effects, such as the nuclear modification of the Parton Distribution Function, are also expected to have a significant effect on the measured distribution.

\end{abstract}
\end{titlepage}
\setcounter{page}{2}

%
%

\section{Introduction}

In ultra-relativistic heavy-ion collisions at the Relativistic Heavy-Ion Collider (RHIC) and at the Large Hadron Collider (LHC),
strongly-interacting matter characterized by high energy density and temperature is
produced \cite{ARSENE20051, ADCOX2005184, BACK200528, ADAMS2005102, Aamodt:2010pa, Aamodt:2010jd}.
Under these conditions, the formation of a deconfined state of quarks and gluons,
called Quark-Gluon Plasma (QGP), is predicted by Quantum ChromoDynamic (QCD) calculations on
the lattice~\cite{pQCD1,pQCD2,pQCD3,pQCD4,pQCD5}.
The production of heavy quarks, i.e. charm (c) and beauty (b), takes place via initial partonic
scattering processes with large momentum transfer (hard scattering) on a timescale
of $\hbar/(2\,m_{\rm c, b}\,c^2$), where $m$ is the mass of the quark. This timescale
(e.g. $\approx$ 0.08 fm/$c$ for charm) is smaller than the QGP thermalization time
($\approx$ 0.6--1\,fm/$c$~\cite{Liu:2012ax}). Additional thermal production as well as
annihilation rates of charm and beauty quarks in the strongly interacting medium are expected to be small in
Pb--Pb collisions even at LHC energies~\cite{Averbeck:2013oga,0954-3899-43-9-093002,Annihalition}.
Consequently, charm and beauty quarks are ideal probes to investigate the properties of the QGP,
since they experience the full evolution of the strongly interacting medium produced in
high-energy heavy-ion collisions.

In order to exploit the sensitivity of heavy-flavour observables to medium effects 
a precise reference where such effects are not expected is needed and it is provided by pp collisions. In pp collisions, 
heavy-quark production can be described theoretically via perturbative QCD calculations
over the full quark momentum range, while such a description does not hold for gluon and light-quark production~\cite{Averbeck:2013oga}. Therefore, measurements of heavy-flavour production
cross sections in pp collisions are used to test perturbative QCD calculations and provide the necessary experimental reference for heavy-ion collisions.

The modification of the $p_{\rm T}$-differential yield in heavy-ion collisions with 
respect to pp collisions at the same centre-of-mass energy is quantified by the nuclear 
modification factor $R_{\rm AA}$, defined as:

\begin{equation}
 R_{\rm AA}(p_{\rm T},y) = \frac{1}{\langle T_{\rm AA} \rangle} \cdot
\frac{{\rm d^{2}}N_{\rm AA}/{\rm d}p_{\rm T}{\rm d}y}{{\rm d^{2}}\sigma_{\rm pp}/{\rm d}p_{\rm T}{\rm d}y}
 \label{raa}
\end{equation}

where d$^2N_{\rm AA}$/d$p_{\rm T}{\rm d}y$ is the yield measured in heavy-ion collisions in a given $p_{\rm T}$ and $y$ interval, and 
d$^2\sigma_{\rm pp}$/d$p_{\rm T}{\rm d}y$ is the corresponding production cross section in pp collisions. 
The average nuclear overlap function, $\langle$$T_{\rm AA}$$\rangle$, is given by the ratio of the average number of 
binary nucleon-nucleon collisions in a centrality class and the inelastic nucleon-nucleon cross section, and it is 
determined via Glauber model calculations~\cite{Glaubermodela,Glaubermodelb}. In the absence of medium effects, $R_{\rm AA}$ 
is expected to be unity for hard probes such as charm and beauty production. 

For momenta larger than the masses of charm and beauty quarks, the dominant medium effect is the partonic energy loss 
via radiative~\cite{Radiativea, Radiativeb} and collisional processes~\cite{Colla, Collb, Collc} when heavy quarks propagate 
through the QGP. These processes are expected to cause a shift of the partonic momentum distribution towards lower momenta 
and, therefore, to lead to a suppression of the yield of heavy-flavour hadrons and their decay products at high 
$p_{\rm T}$ ($\gtrsim~ 2$~GeV/$c$) and, consequently, to $R_{\rm AA} < 1$. In the absence of further processes that modify 
the total charm and/or beauty production cross section or the fragmentation/hadronization of heavy
quarks, $R_{\rm AA}$ is expected to increase again towards low $p_{\rm T}$ to compensate the suppression at high $p_{\rm T}$ 
and, therefore, conserve the binary collision scaling. At RHIC, such a rise was observed by the PHENIX and STAR experiments for 
leptons from semileptonic heavy-flavour hadron decays in Au-Au and Cu-Cu collisions at 
\energy~=~200~GeV~\cite{STARRaaeAuAu,PHENIXRaaeAuAu,PHENIXRaaeAuAubis,Adare:2013yxp}.
The STAR Collaboration also measured the $R_{\rm AA}$ of D$^{0}$ mesons in Au-Au collisions for $p_{\rm T} < 8$~GeV/$c$~\cite{STARRaaD0AuAu}.

The interaction of charm and, to a lesser extent, beauty quarks 
of low transverse momentum with the medium may lead to the participation of heavy quarks in the collective expansion 
of the hot and dense system~\cite{Adam:2016ssk, Abelev:2013lca} and, eventually, to a partial or complete thermalization 
of heavy quarks in the system~\cite{vanHees:2005wb}. Moreover, while in pp collisions charm and 
beauty quarks hadronize via fragmentation, in heavy-ion collisions a competing hadronization 
mechanism through the coalescence with other quarks from the medium could become relevant 
and modify the phase-space distribution of heavy-flavour hadrons up to transverse momenta 
of a few GeV/$c$~\cite{Recombinationa, Recombinationb, Andronic:2015wma}.
Finally, initial-state effects due the presence of a heavy nucleus in the collision
system can play a role. At low Bjorken-$x$ (below $10^{-2}$) the parton densities in nucleons 
bounded in nuclei are reduced with respect to those in free nucleons. This so-called "shadowing" 
leads to a reduction of heavy-flavour production, becoming more pronounced with decreasing
$p_{\rm T}$~\cite{Eskola:2009uj}. In addition, at lower collision energies, momentum ($k_{\rm T}$)
broadening leads to an enhancement of $R_{\rm AA}$ at intermediate $p_{\rm T}$, the so-called
Cronin effect~\cite{PhysRevLett.88.232303}.

At the LHC, open heavy-flavour production was measured in Pb--Pb collisions via exclusive
hadron decays of prompt D and B mesons and via leptons from heavy-flavour hadron 
decays~\cite{Sirunyan:2017xss, Sirunyan:2017oug, Adam:2016wyz, ALICEDRPbPb,ALICEDRPbPbcent,ALICEmuRPbPb,Adam:2016khe,Adam:2015sza}.
At high $p_{\rm T}$ ($\gtrsim~3$~GeV/$c$), a substantial suppression with respect to the scaled 
reference cross section from pp collisions is observed with $R_{\rm AA}$ values similar to 
those measured at RHIC. At lower $p_{\rm T}$, the $R_{\rm AA}$ of prompt D mesons stays below 
unity down to transverse momenta as low as 1~GeV/$c$, in contrast to corresponding measurements 
at RHIC where $R_{\rm AA}$ reaches a maximum value of $\approx$1.5 at $p_{\rm T} \approx$ 1--2 GeV/$c$. 
The different patterns observed at the LHC and at RHIC could be due to differences in the initial 
momentum distributions of heavy quarks, the magnitude of parton energy loss in the medium, the 
impact of collective expansion, the relevance of coalescence as a hadronization mechanism, and the 
role of initial-state effects~\cite{Adam:2015sza}.

At the LHC, initial-state effects and their impact on the nuclear modification factor 
are investigated in proton-lead (p--Pb) collisions. The nuclear modification factor $R_{\rm pPb}$
was measured at mid-rapidity for prompt D and B mesons and for electrons from semileptonic
heavy-flavour hadron decays~\cite{Adam:2016wyz, Khachatryan:2015uja, Adam:2015qda, Abelev:2014hha}. The $R_{\rm pPb}$ of electron from heavy-flavour hadron decay was observed to be consistent with unity within uncertainties over the whole $p_{\rm T}$ range of the measurements,
as expected from binary-collision scaling of heavy-flavour production. 

This paper reports on measurements of electrons from semileptonic heavy-flavour hadron 
decays at mid-rapidity ($|y|$ $<$ 0.8) in pp collisions at $\sqrt{s}$ = 2.76 TeV and in Pb--Pb collisions
in the two centrality classes 0--10\% and 20--40\% at \energy~ = 2.76 TeV. The charge averaged $p_{\rm T}$-differential yields, cross sections and the resulting nuclear modification factors are presented.
Applying a data-driven background subtraction technique~\cite{Adam:2015qda} allowed for a reduction of the systematic uncertainties of the pp reference cross section by a factor of about 3 compared to the
previously published reference~\cite{Abelev:2014gla}, which is consistent within uncertainties
with the current measurement.

The results presented in this paper extend the previous measurements~\cite{Adam:2016khe} of electrons from semileptonic 
heavy-flavour hadron decays in Pb--Pb collisions from 3 GeV/$c$ down to 0.5 GeV/$c$ in $p_{\rm T}$.
They complement the measurements of muons from semileptonic
heavy-flavour hadron decays at forward rapidity and of the prompt D mesons at mid-rapidity reported by the ALICE Collaboration~\cite{ALICEmuRPbPb, ALICEDRPbPb}, as well as of muons from semileptonic heavy-flavour hadron decays at mid-rapidity reported by the ATLAS Collaboration \cite{TheATLAScollaboration:2015bqp}. 
The  measured nuclear modification factor $R_{\rm AA}$ is 
compared with model calculations aiming at describing heavy-quark production and energy loss 
in heavy-ion collisions taking into account also initial-state effects. 


\section{Experimental apparatus and data sample} \label{section2}

The ALICE apparatus, described in detail in~\cite{1748-0221-3-08-S08002,Abelev:2014ffa}, consists mainly of a 
central barrel at mid-rapidity ($|\eta|~<$ 0.9) embedded in a solenoidal magnet, and a muon spectrometer at forward 
rapidity ( --4 $<~\eta~<$ --2.5). 
In the following, the subsystems which are used to perform the measurement of electrons from heavy-flavour hadron decays 
are described.

Charged-particle tracks are reconstructed at mid-rapidity ($|\eta|~<$ 0.9) with the Inner Tracking System (ITS) 
and the Time Projection Chamber (TPC). 
The ITS~\cite{1748-0221-5-03-P03003} consists of six cylindrical silicon layers surrounding the beam vacuum pipe. 
The first two layers, made of Silicon Pixel Detectors (SPD) to cope with the high particle density in the
proximity of the interaction point, provide an excellent position resolution of 12 $\mu$$\rm m$ and 100 $\mu$$\rm m$ 
in the r$\varphi$ and the beam direction ($z$-coordinate of the reference system), respectively.
The third and fourth layers consist of Silicon Drift Detectors (SDD), while the two outermost layers are made of Silicon Strip Detectors (SSD). The SDD and SSD layers are also used for charged-particle identification via energy loss (d$E$/d$x$) measurements. 

The TPC~\cite{Alme:2010ke} is the main tracking detector in the central barrel and provides 
a charged-particle momentum measurement together with excellent two-track separation and particle identification via d$E$/d$x$ determination. 

The Time-Of-Flight (TOF) detector~\cite{tofperf} provides the measurement of the time-of-flight for charged particles 
from the interaction point up to the detector radius of 3.8 m, with an overall resolution of about 80 ps. 
The measured time-of-flight of electrons is well separated from that of kaons and protons up to 
$p_{\rm T}$ $\simeq$ 2.5 GeV/$c$ and $p_{\rm T}$ $\simeq$ 4 GeV/$c$, respectively.

The V0 detectors~\cite{vZero} consist of two arrays of 32 scintillator tiles covering the pseudorapidity ranges 
2.8 $<$ $\eta$  $<$ 5.1 (V0A) and \mbox{$-$3.7 $<$ $\eta$ $<$ $-$1.7} (V0C), respectively, and are used for 
triggering and for centrality estimation. 
The latter is performed through a Glauber Monte Carlo (MC) fit of the signal amplitude in the two scintillator 
detectors~\cite{centralitypaper, Loizides:2014vua, Alver:2008aq}.
Together with the Zero Degree Calorimeters (ZDC)~\cite{Arnaldi:409740}, located on both sides of the interaction 
point at z $\approx$ $\pm$114\,m, they are used offline for event selection.

The pp results presented in this paper are based on the same minimum-bias (MB) data sample recorded at 
$\sqrt{s}$~=~2.76\,TeV as the previously published result \cite{Abelev:2014gla}. The MB trigger required 
at least one hit in the SPD or a signal (above threshold) in either of the two V0 arrays, in temporal coincidence 
with a signal from the beam position monitors~\cite{Abelev:2014ffa}. Pile-up events are identified and rejected 
using the SPD~\cite{Abelev:2012xe,Abelev:2014gla}, and they amount to about 0.7\% of all events. During the 
pp run at 2.76 TeV, the information from the SDD was read out only for a fraction of the recorded events to maximize 
the data acquisition speed. For the current analysis all events have been reconstructed without the SDD 
information in order to obtain a homogeneous sample over the full statistics.
 
For the Pb--Pb analysis, the same data sample recorded at \energy~ = 2.76\,TeV was used as for previous 
publications~\cite{Adam:2016khe, Adam:2016ssk}. The events were 
collected with a MB interaction trigger using information from the coincidence of signals between the 
V0A and V0C detectors. Central and semi-central Pb--Pb collisions were selected online by applying different thresholds 
on the V0 signal amplitudes resulting in central (0--10\%) and semi-central (10--50\%) 
trigger classes~\cite{Abelev:2014ffa}. 
Events affected by 
pile-up from different bunch crossings have been rejected offline~\cite{Adam:2016ssk}. This selection 
removes up to 5\% of the total number of events depending on the centrality of the collisions. 

For both collision systems, only events with a reconstructed interaction vertex (primary vertex) within 10~cm 
from the nominal interaction point along the beam direction are used in order to minimize edge effects at the limit 
of the central barrel acceptance.
The number of events analysed after applying the event selection and the corresponding luminosities for the pp and 
the two Pb--Pb centrality classes are listed in Table~\ref{table:event_cutddd}. The values of the average nuclear overlap function for the two Pb--Pb centrality classes are listed as well. These values and the respective uncertainties are updated with respect to the previously published high-$p_{\rm T}$ $R_{\rm AA}$ results~\cite{Adam:2016khe}. More information about the update of the average nuclear overlap function values can be found in \cite{ALICE-PUBLIC-2018-011}.

\begin{table}
\begin{center}
\centering
\begin{tabular}{cccc}
\hline 
Collision system & $N_{\rm events}$ & $\langle T_{\rm AA} \rangle$ (mb$^{-1}$)  \\ 
\hline 
pp & $38.9 \times 10^6$ & --  \\
\hline
Pb--Pb, 0--10\%  & $15.4 \times 10^6$ & 23.37 $\pm$ 0.2 \\
Pb--Pb, 20--40\%  & $8.2 \times 10^6$ & 7.109 $\pm$ 0.15 \\ 
\hline
\end{tabular}
\caption[Event selection]{Number of events for the pp collisions and the two Pb--Pb centrality classes after applying the event selection. In the right column the average nuclear overlap function is reported for the Pb--Pb samples \cite{ALICE-PUBLIC-2018-011}.}
\label{table:event_cutddd}
\end{center}
\end{table} 

\section{Data analysis} \label{section3}

The $p_{\rm T}$-differential yield of electrons from semileptonic heavy-flavour hadron decays is computed by 
measuring the inclusive electron yield and subtracting the contribution of electrons that do not originate 
from open heavy-flavour hadron decays.
In the following, the inclusive electron identification strategy and the subtraction of electrons 
originating from background sources are described for the analysis of pp and Pb--Pb collisions.

\subsection{Track selection and electron identification}

Candidate electrons tracks are required to fulfil the criteria summarized in Table~\ref{table:track_cuts}, similarly to what was done in Refs.~\cite{Adam:2016ssk, Abelev:2014gla}, in order to select good quality tracks. The rapidity range used in the analyses is restricted to $|y| < 0.8$ to exclude the edges of the detectors, where the systematic uncertainties related to particle identification increase. 

 \begin{table}[ht!]
\begin{center}
\centering
   \begin{tabular}{ccc}
\hline 
Data Sample  & Pb--Pb & pp \\
\hline
$p_{\rm T}$ range (GeV/$c$) & 0.5--3 & 0.5--3 \\
$|y|$ & $<$ 0.8 & $<$ 0.8 \\
\hline
Number of TPC clusters &  $\ge$ 100 & $\ge$ 110 \\
Number of TPC clusters in d$E$/d$x$ calculation &  $\ge$ 90 & $\ge$ 80 \\
Ratio of found TPC clusters over findable &  $>$ 0.6 & $>$ 0.6\\
$\chi^2$/clusters of the momentum fit in the TPC & $<$  3.5 & $<$  4 \\
DCA$_{xy}$  & $<$ 2.4 cm & $<$ 1 cm\\
DCA$_{z}$ & $<$ 3.2 cm & $<$ 2 cm\\
Number of ITS hits & $\ge$ 5  & $\ge$ 3  \\
Number of hits in the SPD layers & 2 &  2  \\
\hline 
\end{tabular}
\caption{Track selection criteria used in the analyses. DCA is an abbreviation for the distance of closest approach of a track to the primary vertex.}
\label{table:track_cuts}
\end{center}
\end{table}

The electron identification is mainly based on the measurement of the specific ionization energy loss in the TPC (d$E$/d$x$), 
similarly to the procedure followed in Refs.~\cite{Adam:2016ssk, Abelev:2014gla}. The discriminant variable is the deviation of d$E$/d$x$ from the parametrized electron Bethe-Bloch~\cite{BetheBloch} expectation value, 
expressed in units of the d$E$/d$x$ resolution, $n_{\sigma}^{\rm{TPC}}$~\cite{Abelev:2014ffa}. 

In order to reduce the hadron contamination in Pb--Pb collisions, tracks with a time-of-flight differing from the expected value for electrons ($n_{\sigma}^{\rm{TOF}}$) by twice the TOF resolution or more are rejected. In pp collisions, 
a $|n_{\sigma}^{\rm{TOF}}| \geq 3$ rejection is applied due to the smaller hadron contamination. 

In Pb--Pb collisions, in addition, the d$E$/d$x$ in the ITS is used to further reject hadrons. To guarantee a good 
Particle IDentification (PID) based on the d$E$/d$x$ in the ITS, tracks are required to have at least three out of 
the four possible hits in the external layers of the ITS (SDD and SSD), which can provide d$E$/d$x$ 
measurements. 
Table~\ref{table:pid_cuts} summarizes the PID selection criteria for electron identification.

\begin{table}[h!]
\begin{center}
\centering
   \begin{tabular}{ccccc}
\hline
&$p_{\rm T}$ range (GeV/$c$) & TPC d$E$/d$x$  & ITS d$E$/d$x$  & TOF compatibility  \\
& & selection &selection  & with $e$ hypothesis   \\
\hline
pp&\,\, 0.5--3 & $-$1 $<$ $n_{\sigma}^{\rm{TPC}}$ $<$ 3 & -- & $|$$n_{\sigma}^{\rm{TOF}}$$|$ $<$ 3  \\
\hline 
Pb--Pb&\,\, 0.5--1.5 & $-$1 $<$ $n_{\sigma}^{\rm{TPC}}$ $<$ 3 & $|$$n_{\sigma}^{\rm{ITS}}$$|$ $<$ 1 & $|$$n_{\sigma}^{\rm{TOF}}$$|$ $<$ 2  \\
&1.5--3 & \, \, 0 $<$ $n_{\sigma}^{\rm{TPC}}$ $<$ 3 & $|$$n_{\sigma}^{\rm{ITS}}$$|$ $<$ 2 & $|$$n_{\sigma}^{\rm{TOF}}$$|$ $<$ 2  \\
\hline
\end{tabular}
\caption{Electron identification criteria used in the analyses (see text for more details).}
\label{table:pid_cuts}
\end{center}
\end{table} 

The remaining hadron contamination 
is estimated by fitting in momentum slices the TPC d$E$/d$x$ distribution after the TOF (and ITS) PID 
selections~\cite{Abelev:2012xe, Adam:2016ssk}. The hadron contamination is negligible at the lowest $p_{\rm T}$
and it increases with $p_{\rm T}$, reaching about 5\% at $p_{\rm T}$ = 3 GeV/$c$ in Pb--Pb collisions and 
about 1\% in pp collisions, with negligible dependence on centrality and pseudorapidity. In both collision systems 
the hadron contamination is subtracted statistically from the inclusive electron candidate yield.

\subsection{Subtraction of electrons from non heavy-flavour sources} \label{NHFE}

The raw inclusive sample of electron candidates ($p_{\rm T}$ $<$ 3 GeV/$c$) consists of the signal, i.e. the 
electrons from semileptonic heavy-flavour hadron decays, and four background components:

\begin{enumerate}
\item photonic electrons from Dalitz decays of light neutral mesons (predominantly $\pi^{0}$ and $\eta$ mesons) and the conversion of their decay photons in 
      the detector material, as well as from prompt virtual and real photons from thermal and hard scattering 
      processes; 
\item electrons from weak ${\rm K^{0/\pm}} \rightarrow {\rm e}^{\pm}\pi^{\mp/0}\,\overset{\scriptscriptstyle(-)}{\nu_{e}}$ (${\rm K}_{{\rm e}3}$) decays;
\item dielectron decays of quarkonia;
\item dielectron decays of light vector mesons.
\end{enumerate}

The photonic-electron tagging method~\cite{Adamczyk:2014yew, Adam:2015qda} is adopted for the subtraction of the first and main background component. For $p_{\rm T}$ $<$ 1.5 GeV/$c$ the inclusive electron yield is largely dominated by the contribution of photonic electrons. 
The ratio of the signal to the photonic electron background is measured to be 0.2 at $p_{\rm T}$ = 0.5 GeV/$c$ and it is observed to increase reaching a value of 3 at $p_{\rm T}$ = 3 GeV/$c$ ~\cite{Adam:2016ssk}.
Photonic electrons originate from electron-positron pairs with a small invariant mass ($m_{{\rm e}^{+}{\rm e}^{-}}$).  
They are tagged by pairing an electron (positron) track with opposite charge tracks identified as 
positrons (electrons) from the same event. The latter are called associated electrons in the following and they are selected with less stringent requirements listed in Table~\ref{table:track_cutsasso}. The combinatorial background from uncorrelated electron-positron pairs is subtracted using as a proxy the like-sign 
invariant mass distribution in the same invariant mass interval. 
A selection on the pair invariant mass is applied as listed in Table~\ref{table:track_cutsasso}.

\begin{table}[ht!]
\begin{center}
\centering
   \begin{tabular}{ccc}
\hline
Associated electron &  Pb--Pb & pp   \\
\hline 
  $p_{\rm T}$ (GeV/$c$)& $>$ 0.15 & $>$ 0.1\\
  $|y|$ & $<$ 0.9 & $<$ 0.8 \\
Number of TPC clusters &  $\ge$ 80& $\ge$ 60\\
Number of ITS hits &  $\ge$ 2&$\ge$ 2 \\
DCA$_{xy}$ & $<$ 2.4\,cm& $<$1 cm\\
DCA$_{z}$ & $<$ 3.2\,cm& $<$ 2 cm \\
TPC d$E$/d$x$  &  $|n_{\sigma}^{\rm{TPC}}|$ $<$ 3  &  $|n_{\sigma}^{\rm{TPC}}|$ $<$ 3 \\

\hline
Electron-positron pair  &   &\\
\hline
$m_{{\rm e}^{+}{\rm e}^{-}}$ (MeV/$c^2$)& $<$ 70& $<$ 140 \\
\hline
\end{tabular}
\caption{Selection criteria for tagging photonic electrons in Pb--Pb and pp collisions.} 
\label{table:track_cutsasso}
\end{center}
\end{table} 

Due to detector acceptance and inefficiencies and because of the decay kinematics, not all photonic electrons in the 
inclusive electron sample are tagged with this method. Therefore, the raw yield of tagged photonic electrons is corrected 
for the efficiency to find the associated electron (positron), hereafter called tagging efficiency.
This efficiency is estimated with Monte Carlo (MC) simulations. In particular, HIJING v1.383~\cite{Hijing:ref} was used 
to simulate Pb--Pb collisions, while the PYTHIA 6 (Perugia 2011 tune)~\cite{Sjostrand:2006za} event generator was used 
for the simulation of pp events.
The transport of particles in the detector is performed using GEANT3~\cite{Brun:1994aa}. In both analyses, 
the generated $\pi^{0}$ $p_{\rm T}$ distributions in MC are weighted so as to match the measured neutral pion $p_{\rm T}$ 
spectra~\cite{NeutralPionSpectra,ChargedPionSpectra}. In the pp analysis, the $\eta$ $p_{\rm T}$ spectra 
are weighted using the corresponding measurement~\cite{Acharya:2017hyu}, while for Pb--Pb collisions the $\eta$ weights 
are determined via $m_{\rm T}$-scaling of the measured $\pi^0$ $p_\text{T}$ spectra~\cite{Albrecht:1995ug, Khandai:2011cf}. 
The resulting $\eta/\pi^{0}$ ratios agree within uncertainties with the ratios measured by ALICE in 0-10\% and 20-50\%
central Pb--Pb collisions at \energy~ = 2.76~TeV~\cite{Acharya:2018yhg}.
The photonic electron tagging efficiency increases with the electron $p_{\rm T}$, starting from a value of $\approx$40\% 
($\approx$30\%) at $p_{\rm T}$ = 0.5 GeV/$c$ and reaching a value of $\approx$70\% ($\approx$60\%) at $p_{\rm T}$ = 3 GeV/$c$ 
for pp (Pb--Pb) collisions.   

The background contribution of non-photonic electrons from ${\rm K}_{{\rm e}3}$ decays and the dielectron decay of
${\rm J}/\psi$ mesons is subtracted from the fully corrected and normalized electron yield using the so-called
cocktail approach in both pp and Pb--Pb collisions~\cite{Adare:2010de, Abelev:2014gla, Abelev:2012xe,  Adam:2015qda}.
Due to the requirement of hits in both pixel layers, the relative contribution from ${\rm K}_{{\rm e}3}$ decays to
the electron background is small and it decreases with $p_{\rm T}$, with a maximum of about 0.5\% at
$p_{\rm T}$ = 0.5 GeV/$c$ for both the collision systems. For pp collisions, the contribution of electrons from 
${\rm J}/\psi$ decays is calculated based on a phenomenological interpolation of the ${\rm J}/\psi$ production 
cross sections measured at various values of $\sqrt{s}$ as described in~\cite{Bossu:2011qe}, and as done in a 
previous analysis ~\cite{Abelev:2014gla}.
For Pb--Pb collisions, the $p_{\rm T}$-differential ${\rm J}/\psi$ yield is calculated by multiplying this
reference ${\rm J}/\psi$ cross section in pp collisions with $\langle T_{\rm AA} \rangle$ and the measured
nuclear modification factor in Pb--Pb collisions~\cite{Adam:2015rba, Adam:2016rdg}.
The contribution of electrons from ${\rm J}/\psi$ decays is maximal in the interval $2.0 < p_\text{T} < 3.0$~GeV/$c$, with a value of $\approx 3$\% in pp collisions and of $\approx 5$\% in central Pb--Pb collisions. At higher $p_{\rm T}$ and in less central Pb--Pb collisions the background from ${\rm J}/\psi$ decays decreases. At lower $p_{\rm T}$ it is negligible.  The background from dielectron decays of light vector mesons and other quarkonium states as well as from Dalitz decays of higher mass mesons 
($\omega$, $\eta'$, $\phi$) is negligible as discussed in Ref.~\cite{Adam:2016ssk}.

\subsection{Correction and normalisation}

After the statistical subtraction of the hadron contamination and the background from photonic electrons, the raw yield of electrons and positrons is divided by the number of events analysed ($N^{MB}_\text{ev}$), by the value of $p_{\rm T}$ at the centre of each bin and its width $\Delta p_{\rm T}$, by the width $\Delta y$ of the covered rapidity interval, by the geometrical acceptance ($\epsilon^\text{geo}$) times the reconstruction ($\epsilon^\text{reco}$) and PID efficiencies ($\epsilon^\text{eID}$) and a factor of two to obtain the charge averaged invariant differential yield

\begin{equation}
 \frac{1}{2\pi p_\text{T}} \frac{\text{d}^2 N^{e^\pm}}{\text{d}p_\text{T} \text{d}y} = \frac{1}{2} \frac{1}{2\pi\,p_\text{T,centre}} \frac{1}{N^{MB}_\text{ev}} \frac{1}{\Delta y \Delta p_\text{T}} \frac{N^{e^\pm}_\text{raw}(p_\text{T})}{(\epsilon^\text{geo} \times \epsilon^\text{reco} \times \epsilon^\text{eID})}. 
 \label{norm}
\end{equation}

The invariant production cross section in pp collisions is obtained by further multiplying with the minimum-bias trigger cross section for pp collisions at $\sqrt{s} = 2.76$~TeV, $\sigma_\text{MB}$ = ($55.4 \pm 1.0$)~mb~\cite{Abelev:2012sea}.

The efficiencies are determined using dedicated MC simulations. The reconstruction efficiencies are computed using a heavy-flavour enriched PYTHIA 6~\cite{Sjostrand:2006za} MC sample in which each simulated pp event contains a $c\bar{c}$ or $b\bar{b}$ pair, and heavy-flavour hadrons are forced to decay semi-electronically.
In the MC production used for the Pb--Pb analysis the underlying events are simulated using the HIJING v1.383 generator~\cite{Hijing:ref} and the heavy-flavour signal from the PYTHIA 6 generator is added.
Out of all produced particles in these PYTHIA pp events, only the heavy-flavour decay products are kept and transported through the detector together with the particles produced with HIJING.
In order to better reproduce  the experimental conditions for the detector occupancy, the number of heavy quarks injected into each HIJING event is adjusted according to the Pb--Pb collision centrality.
In Pb--Pb collisions, the bin-wise total reconstruction efficiencies 
($\epsilon^\text{geo} \times \epsilon^\text{reco} \times \epsilon^\text{eID}$) do not show any significant $p_{\rm T}$ dependence 
and are about 8\% (9\%) in the 0--10\% (20--40\%) centrality class. Due to the less stringent selections applied for pp collisions, 
the total electron reconstruction efficiency reaches a value of about 27\% at $p_{\rm T}$ = 3 GeV/$c$ in this case.
Finally, the remaining background contributions from weak ${\rm K}_{{\rm e}3}$ decays and dielectron decays
of ${\rm J}/\psi$ mesons are subtracted from the fully corrected cross section (yield) for pp (Pb--Pb) collisions.

\subsection{Systematic uncertainties} 

The overall systematic uncertainty on the $p_{\rm T}$ spectra is calculated summing in quadrature the different uncorrelated contributions,
which are summarised in Table~\ref{syst} and discussed in the following.

The systematic uncertainties arising from the residual discrepancy between MC used to determine the total reconstruction efficiency and data
is estimated by systematically varying the track selection and PID requirements around the default values chosen in the analysis. The systematic
uncertainties are determined as the root mean squared (RMS) of the distribution of the resulting corrected yields (or cross sections in pp)
obtained for different selections in each $p_{\rm T}$ interval, considering also shifts of the mean value with respect to the default selections. In the Pb--Pb analysis, this contribution is about 6\% at low $p_{\rm T}$
($p_{\rm T} < 1$~GeV/$c$), and it decreases with increasing $p_{\rm T}$ reaching about 3\% at the highest $p_{\rm T}$.
In the pp case this contribution is about 4\% without $p_{\rm T}$ dependence.

In the pp analysis, a systematic uncertainty of about 2\% (3\%) is assigned due to the incomplete knowledge of the efficiency in matching
tracks reconstructed in the ITS and TPC (TPC and TOF)~\cite{Abelev:2012xe,Abelev:2014gla}. In Pb--Pb collisions, the uncertainty assigned
on the measurements coming from the track-reconstruction procedure amounts to 5\% for single tracks~\cite{Abelev:2012hxa}.

The solenoid polarity was changed during the Pb--Pb data taking period. From the comparison of the fully corrected spectra of electrons
from semileptonic heavy-flavour hadron decays measured in events with the magnetic field oriented in the two opposite directions,
a 2\% systematic uncertainty is assigned for $p_{\rm T}~\leq$ 1.25 GeV/$c$. To ensure that the results are not biased by tracks detected
at the edges of the detector, where the efficiencies are more difficult to be calculated, the measurements were re-done restricting the
rapidity window for the electrons down to $|y|~<~0.5$. In addition, possible biases in the efficiency determination are checked by
performing the analyses only in the positive or the negative rapidity region. A 5\% systematic uncertainty has been estimated for
$p_{\rm T} < 1.5$~GeV/$c$ in both pp and Pb--Pb collisions.

The systematic uncertainty arising from the photonic-electron subtraction technique is estimated similarly as the RMS of the distribution
of yields obtained by varying the selection criteria listed in Table~\ref{table:track_cutsasso}. In the Pb--Pb analysis, because of the
large combinatorial background of random pairs, this systematic uncertainty is of the order of $\pm$30\% in the 0--10\% most-central
collisions and $\pm$18\% in the centrality class 20--40\% for the $p_{\rm T}$ interval 0.5--0.7 GeV/$c$. It is observed to decrease with
increasing $p_{\rm T}$ reaching 2\% for $p_{\rm T}$ = 2 GeV/$c$, where the contribution of background electrons starts to become negligible.
In pp collisions, the uncertainty arising from the photonic-electron subtraction is estimated to be about 3\% with no $p_{\rm T}$ dependence.
In addition, the dependence of the photonic-electron tagging efficiency on the spectral shape of the background sources is taken into account
by recalculating the efficiency for different $\pi^{0}$ and $\eta$ $p_{\rm T}$ spectra. The variation of the neutral-meson spectra is obtained
by parameterising the measured spectra considering their systematic uncertainties. In particular, the measured yields at the lowest
transverse momenta are shifted up by their systematic uncertainties and the yields at the highest transverse momenta are shifted down, and vice versa.
The resulting systematic uncertainty on the spectra of electrons from semileptonic heavy-flavour hadron decays is 1\% for $p_{\rm T}~\leq$ 0.9 GeV/$c$
in Pb--Pb collisions. In pp collisions, the systematic uncertainty is about 5\% in the $p_\text{T}$ interval 0.5--0.7 GeV/$c$, 2\% in 0.7--0.9 GeV/$c$,
1\% in 0.9--1.5 GeV/$c$ and negligible for higher $p_\text{T}$. It is worth noting that replacing the previous approach to determine the photonic
background via a cocktail calculation of the known sources~\cite{Abelev:2014gla} by an actual measurement of this background component resulted
in a reduction of the related systematic uncertainties of the pp reference cross section by a factor of about 3.

In order to further test the robustness of the photonic-electron tagging, the number of clusters required for electron candidates 
in the SPD has been released to
a single hit in any of the two layers, increasing in this way the fraction of electrons coming from photon conversions in the detector material.
In the pp analysis, a contribution to the systematic uncertainties of about 20\% in the $p_\text{T}$ interval 0.5--0.7 GeV/$c$ and 5\% up to
$p_\text{T}~=$ 1.3 GeV/$c$ is assigned, while for higher $p_{\rm T}$ this uncertainty is estimated to be negligible.
In the Pb--Pb case the systematic uncertainty is 3\% with no $p_{\rm T}$ and centrality dependence. This systematic uncertainty is significantly larger
for the pp sample because of the specific detector configuration. Due to the lack of the SDD detector information at track reconstruction level,
only a maximum of four hits in the ITS can be expected instead of the usual six. Therefore, this sample is potentially affected by a higher fraction
of badly reconstructed tracks, particularly at the lowest transverse momenta. In addition to releasing the condition on the SPD layers, the
systematic uncertainty in the pp case has been determined by comparing the measurement obtained from the analysis of a sub-set of events
where all six ITS layers are used for the track reconstruction.

The subtraction of the background electron contribution from the $\text{J}/\psi$ and $\text{K}_{e3}$ decays is affected by the uncertainty
on the input distribution used for the cocktail calculation. This results in an uncertainty of 4\% and 2\%  in the lowest $p_\text{T}$ interval
in pp and in Pb--Pb collisions, respectively. While for pp collisions this contribution is negligible at higher $p_\text{T}$, for Pb--Pb collisions
it decreases slowly with increasing $p_\text{T}$, reaching a minimum of 1\% at $p_\text{T} = 1.5$~GeV/$c$ before increasing again to 4\%
at $p_\text{T} = 3$~GeV/$c$ due to the growing contribution from $\text{J}/\psi$ decays.

Events with a primary vertex reconstructed using charged-particle tracks are used. For the pp analysis, the resolution of the vertex position
is affected by the absence of the SDD information and by the lower multiplicity of tracks compared to the Pb--Pb case. The associated
uncertainty of 3\% is estimated by comparing the cross sections measured from events where the vertex was determined either with
charged-particle tracks or with the SPD information only.

 \begin{table}[ht!]
\begin{center}
\centering
   \begin{tabular}{lcccccc}
\hline
Collision system & \multicolumn{2}{c}{Pb--Pb (0-10\%)} & \multicolumn{2}{c}{Pb--Pb (20-40\%)} & \multicolumn{2}{c}{pp} \\
\hline
$p_{\rm T}$ interval (GeV/$c$) & 0.5--0.7 & 2--3 & 0.5--0.7 & 2--3 & 0.5--0.7 & 2--3 \\
\hline
Electron candidate selection & 6\% & 3\% & 6\% & 3\% & \multicolumn{2}{c}{4\%} \\
Photonic electron subtraction & 30\% &  2\%  & 18\% &  2\% & \multicolumn{2}{c}{3\%} \\
$\pi^{0}$ and $\eta$ Weights & 1\% & - & 1\% & - & 5\% &  - \\
SPD requirement & \multicolumn{2}{c}{3\%}  & \multicolumn{2}{c}{3\%}  & 20\% &  - \\
Track matching  & \multicolumn{2}{c}{5\%}  & \multicolumn{2}{c}{5\%}  &  \multicolumn{2}{c}{4\%} \\
Magnet polarity & 2\% & - & 2\% & - & \multicolumn{2}{c}{-} \\
Rapidity range & 5\% & - & 5\% & - & 5\% & - \\
Event selection & \multicolumn{2}{c}{-} & \multicolumn{2}{c}{-} & \multicolumn{2}{c}{3\%} \\
Subtraction of $\text{J}/\psi$ and $\text{K}_{e3}$ & 2\% &  4\% & 2\% & 3\% & 4\% & - \\
\hline 
Total systematic uncertainty & 32\% & 8\% & 21\% & 7\% & 23\% & 7\% \\
\hline
\end{tabular}
\caption{Contributions to the systematic uncertainties on the yield of electrons from semileptonic heavy-flavour  
hadron decays, quoted for the lowest and highest $p_{\rm T}$ interval, respectively.}
\label{syst}

\end{center}
\end{table}  



\section{Results} \label{results}

\subsection{$p_\text{T}$-differential invariant cross section in pp collisions} 
\begin{figure}[!ht]
\centering
\includegraphics[scale=0.52]{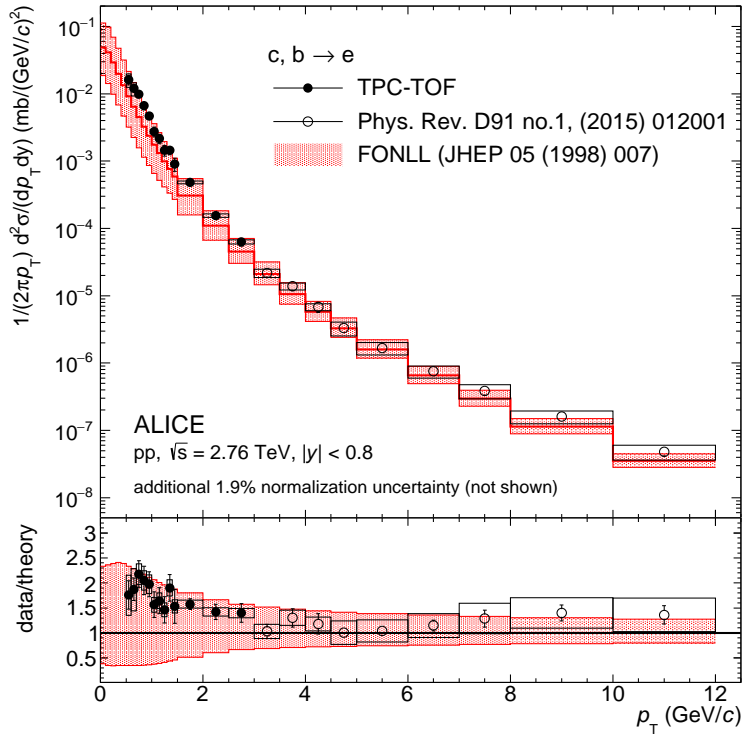}
\caption{The $p_{\rm T}$-differential invariant production cross section for electrons from semileptonic
heavy-flavour hadron decays measured at mid-rapidity in pp collisions at $\sqrt{s} = 2.76$~TeV in 
comparison with FONLL pQCD calculations~\cite{Cacciari:1998it} (upper panel), and the ratio of the data 
to the FONLL calculation (lower panel). Statistical and systematic uncertainties are
shown as vertical bars and boxes, respectively.}
\label{ppspectra}
\end{figure}

The measurement presented in this paper for pp collisions updates the charge averaged $p_\text{T}$-differential 
cross section published previously~\cite{Abelev:2014gla} in the range $p_{\rm T} < 3.0$~GeV/$c$.
The new $p_\text{T}$-differential invariant 
cross section for electrons from semileptonic heavy-flavour hadron decays measured at mid-rapidity 
in pp collisions at $\sqrt{s}$ = 2.76 TeV is shown in \autoref{ppspectra}.  
Results from a previous publication~\cite{Abelev:2014gla} (open circles in \autoref{ppspectra}) are 
plotted together with the new results from the TPC-TOF analysis (filled circles in \autoref{ppspectra}) reported in the
current paper.  
Applying the photonic tagging background subtraction method~\cite{Adam:2015qda} allowed for a reduction of the
systematic uncertainties of the pp reference cross section by a factor of about 3 compared to the
previously published reference~\cite{Abelev:2014gla}, which is consistent within uncertainties
with the current measurement.
The cross section from a pQCD calculation employing the Fixed-Order-Next-to-Leading-Log 
(FONLL) scheme~\cite{Cacciari:1998it} is compared with the data in \autoref{ppspectra}.
The uncertainties of the FONLL calculations (red dashed area) reflect different choices for the charm 
and beauty quark masses, the factorization and renormalization scales as well as from the uncertainty on
the set of parton distribution functions used in the pQCD calculation (CTEQ6.6~\cite{Nadolsky:2008zw}). 
The result from the FONLL calculation is consistent with the 
measured production cross section of electrons from semileptonic heavy-flavour hadron decays. The 
measured cross section is close to the upper edge of the FONLL uncertainty band, as it was observed 
previously in pp collisions at the LHC~\cite{Abelev:2014gla, Abelev:2012xe} and at RHIC, for 
$p_{\rm T} > 1.5$~GeV/$c$~\cite{PHENIXRaaeAuAu, STARRaaeAuAu}, 
as well as in p$\overline{\rm{p}}$ collisions at the Tevatron~\cite{PhysRevLett.91.241804}. 

\subsection{$p_\text{T}$-differential invariant yields in Pb--Pb collisions}
\begin{figure}[!ht]
\centering
\includegraphics[scale=0.55]{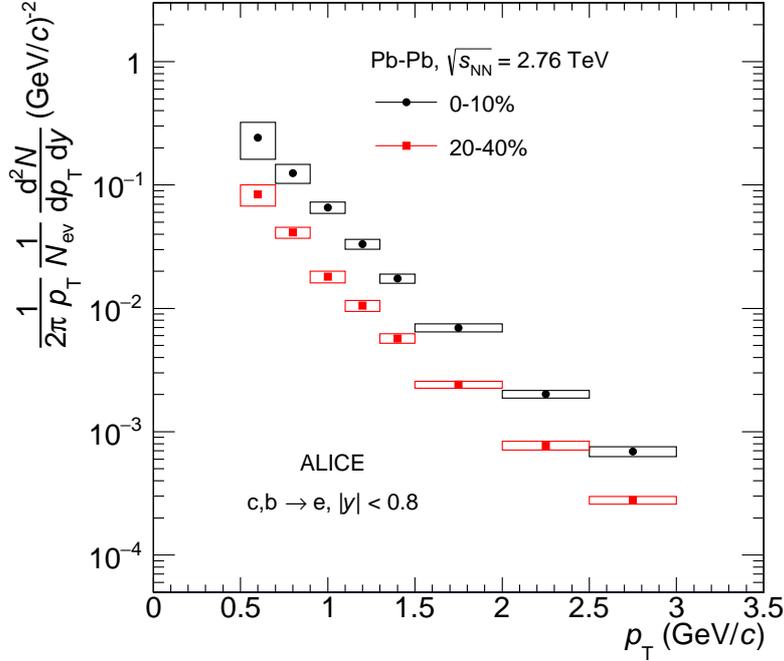}
\caption{The $p_{\rm T}$-differential invariant yields of electrons from semileptonic heavy-flavour 
hadron decays measured at mid-rapidity in 0--10\% and 20--40\% central Pb--Pb collisions at 
$\sqrt{s_{\rm NN}} = 2.76$~TeV. Statistical uncertainties are smaller than the symbol size 
and the systematic uncertainties are shown as boxes.}
\label{spectra}
\end{figure}

The charge averaged $p_\text{T}$-differential invariant yields of electrons and positrons from semileptonic heavy-flavour hadron 
decays measured in the range $0.5 < p_{\rm T} < 3$ GeV/$c$ at mid rapidity in 0--10\% (black circles) and 20--40\% (red squares) central Pb--Pb collisions at 
$\sqrt{s_{\rm NN}} = 2.76$~TeV are depicted in \autoref{spectra}. 

\subsection{Nuclear modification factor $R_{\rm AA}$}

\begin{figure}[!ht]
\centering
\includegraphics[scale=0.83]{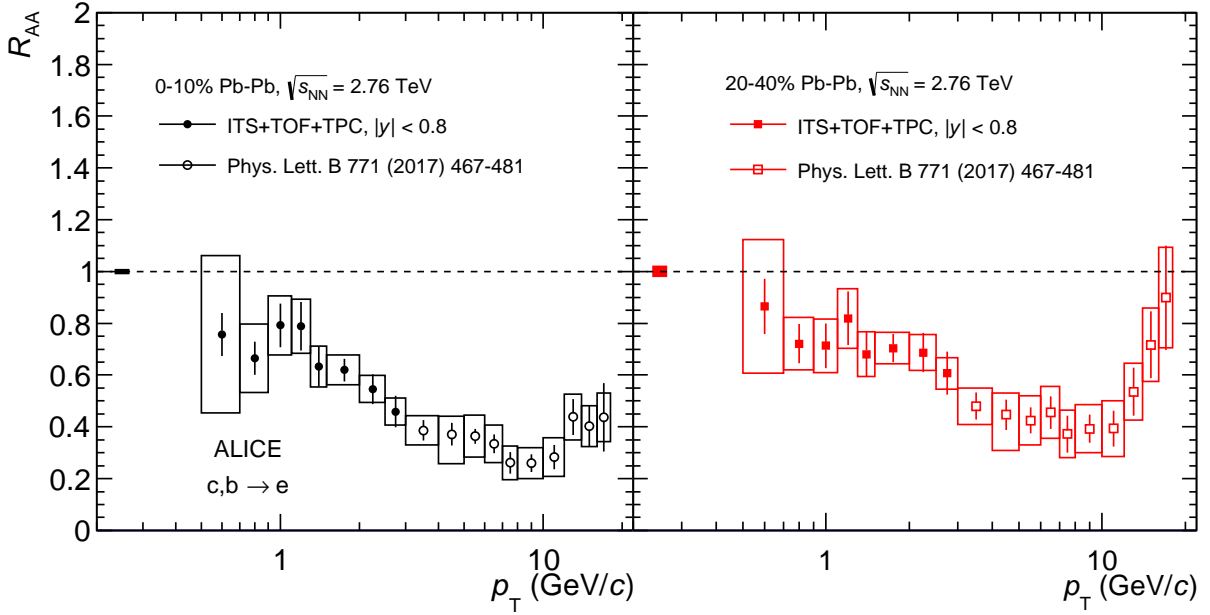}
\caption{Nuclear modification factor $R_{\rm AA}$ for electrons from semileptonic heavy-flavour 
hadron decays at mid-rapidity as a function of $p_{\rm T}$ in 0--10\% (left panel) and 20--40\% 
central (right panel) Pb--Pb collisions at \energy~ = 2.76\,TeV. Error bars (open boxes) represent 
the statistical (systematic) uncertainties. The normalization uncertainties are represented by the 
boxes at $R_{\rm AA} = 1$. The previously published results from~\cite{Adam:2016khe} have been updated using a new glauber model calculation \cite{ALICE-PUBLIC-2018-011}.}
\label{RAA}
\end{figure}

Figure~\ref{RAA} shows the nuclear modification factor of electrons from semileptonic heavy-flavour 
hadron decays at mid-rapidity as a function of $p_{\rm T}$ in Pb--Pb collisions at \energy~= 2.76 TeV 
for the 0--10\% (left  panel) and 20--40\% (right panel) centrality classes.
The low-$p_{\rm T}$ data 
from the current analysis (filled symbols) are shown together with the previously published~\cite{Adam:2016khe} 
high-$p_{\rm T}$ $R_{\rm AA}$ (open symbols). The 20-30\% and 30-40\% centrality intervals from~\cite{Adam:2016khe}, in which electrons were identified using the specific energy loss in the TPC and electromagnetic showers reconstructed in the electromagnetic calorimeter (EMCal) of ALICE, have been combined.
Statistical and systematic uncertainties of the $p_{\rm T}$-differential yields and cross sections in 
Pb--Pb and pp collisions, respectively, are propagated as uncorrelated uncertainties. The 1.9\% 
normalization uncertainty on the pp measurement is included in the systematic uncertainties of the 
invariant cross section, and summed in quadrature with the other systematic uncertainties.
The uncertainties of the average nuclear overlap function $\langle T_{\rm AA} \rangle$ in the 0--10\% 
and 20--40\% centrality classes are represented by the boxes at $R_{\rm AA} = 1$. 
For $p_{\rm T} > 3$~GeV/$c$ the yield of electrons from heavy-flavour hadron decays is suppressed
strongly which was interpreted as due to partonic energy loss in the QGP produced in Pb--Pb 
collisions~\cite{Adam:2016khe}. The current measurement provides an extension of the $p_{\rm T}$ coverage
to lower values, {\it i.e.} from $p_{\rm T} = 3$~GeV/$c$ down to 0.5 GeV/$c$. In this region, the suppression 
of the yield of electrons from heavy-flavour hadron decays is expected to decrease with decreasing $p_{\rm T}$ 
as a consequence of the scaling of the total heavy-flavour yield with the number of binary collisions in 
Pb--Pb collisions. 
This scaling, however can be broken due to the nuclear modification of the parton distribution functions in Pb-nuclei, leading to $p_{\rm T}$-integrated $R_{\rm AA}$ of less than one. Moreover, further modifications of the $p_{\rm T}$ distribution due to the radial flow can also play a role in this region.
The observed $R_{\rm AA}$ in Fig. \ref{RAA} is consistent with the expectation of an increasing $R_{\rm AA}$ with decreasing $p_{\rm T}$, reaching values close to unity within uncertainties.
However the current uncertainties are still too large to quantify the different effects.
Within the current statistical and systematic uncertainties, no significant centrality dependence is
observed in the $p_{\rm T}$-region below 3~GeV/$c$.

\section{Comparison with model calculations} 
\label{modelcomparison}

\begin{figure}[!ht]
\centering
\includegraphics[scale=0.5]{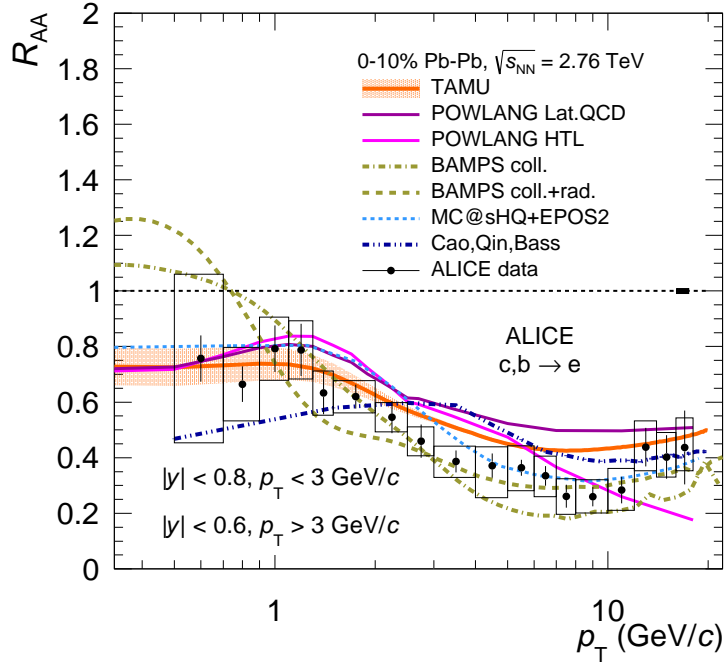}
\caption{$R_{\rm AA}$ of electrons from semileptonic heavy-flavour hadron decays at mid-rapidity as a function of 
         $p_{\rm T}$ in 0--10\% Pb--Pb collisions at \energy~ = 2.76\,TeV compared to model 
         calculations~\cite{Uphoff:2012gb,Uphoff:2013soa,He:2014cla,Alberico:2013bza,Nahrgang:2013xaa, Cao:2013ita}.}
\label{RAAwithmodelfull}
\end{figure}

In \autoref{RAAwithmodelfull} results from model calculations including charm and beauty quark interactions with a QGP medium~\cite{Uphoff:2012gb,Uphoff:2013soa,He:2014cla,Alberico:2013bza,Nahrgang:2013xaa,Cao:2013ita} are compared with 
the measured $R_{\rm AA}$ of electrons from semileptonic heavy-flavour hadron decays for the 10\% most central 
Pb--Pb collisions. The calculations differ in the modelling of the initial conditions, the medium properties, 
the dynamics of the medium evolution, the interactions of charm and beauty quarks with the QGP,
 and in the implementation of hadronisation and hadronic interactions in the 
late stages of the heavy-ion collision. Furthermore, there are differences in the initial $p_{\rm T}$-differential 
heavy-quark production cross section in nucleon-nucleon collisions used as input.
Qualitatively, most models provide a good description of the heavy-flavour $R_{\rm AA}$ measured in the 
most central Pb--Pb collisions as already observed for D mesons~\cite{Adam:2016khe}. 

\begin{figure}[!ht] 
\centering
\includegraphics[scale=0.38]{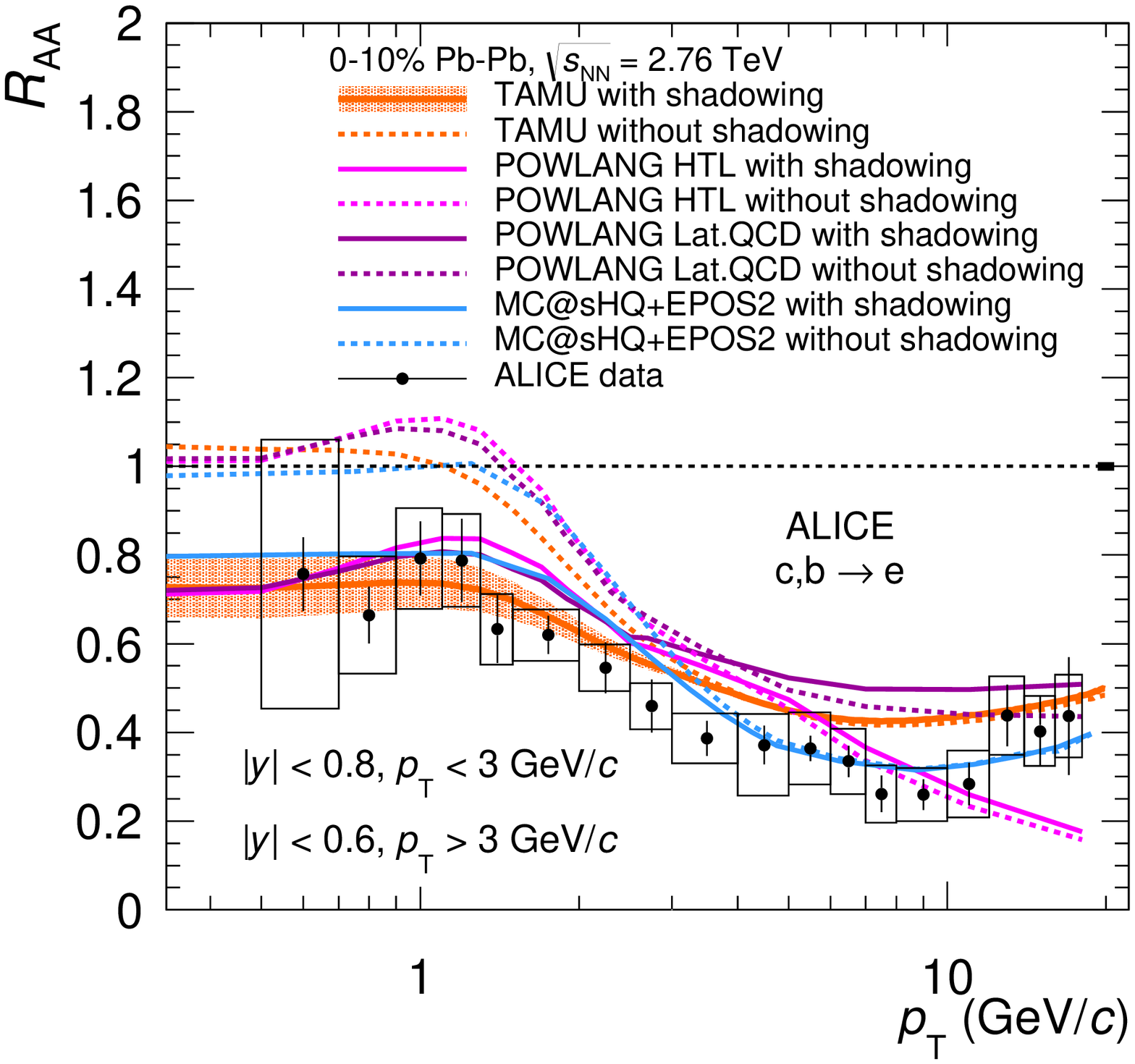}
\includegraphics[scale=0.38]{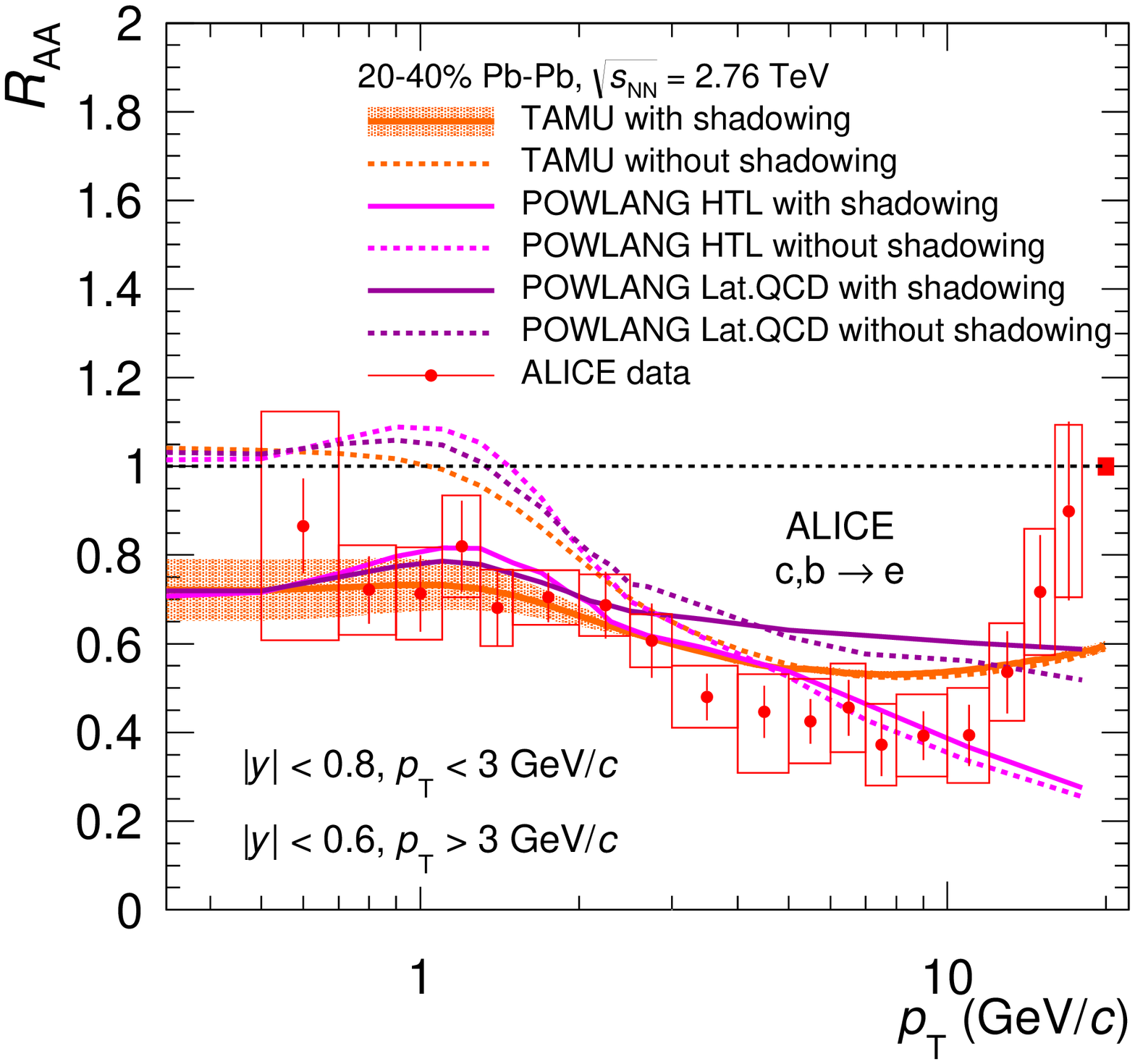}

\caption{$R_{\rm AA}$ of electrons from semileptonic heavy-flavour hadron decays at mid-rapidity as a function of 
         $p_{\rm T}$ in 0--10\% (left) and 20--40\% central (right) Pb--Pb collisions at \energy~ = 2.76\,TeV 
         compared to model calculations~\cite{He:2014cla} with and without EPS09 
         shadowing parameterisations~\cite{Eskola:2009uj}.}
\label{RAAwithTAMU}
\end{figure}

The measurement presented in this paper shows for the first time electrons from heavy-flavour hadron decays in the $p_{\rm T}$ interval below 1 GeV/$c$, where decays of heavy-flavour hadrons down to zero $p_{\rm T}$ contribute.
In this region, the 
nuclear modifications of the PDFs can play a significant 
role~\cite{ ALICEDRPbPb,ALICEDRPbPbcent,ALICEmuRPbPb,Adam:2016khe}. This is addressed in \autoref{RAAwithTAMU}, which compares the measured nuclear modification factor with TAMU, 
POWLANG and MC@sHQ+EPOS2 model calculations with and without the inclusion of the EPS09 shadowing 
parameterisations~\cite{Eskola:2009uj}. The depletion of the parton densities at low $x$, 
resulting in a reduced heavy-flavour production cross section per nucleon-nucleon pair in Pb--Pb 
collisions with respect to bare nucleon--nucleon collisions, leads to a reduction of $R_{\rm AA}$ 
of electrons from heavy-flavour hadron decays at low $p_{\rm T}$. 
Data are better described when the nuclear PDFs are included in the theoretical calculation in both 
centrality intervals. However, the experimental uncertainties are still too large to provide 
quantitative constraints on the nuclear shadowing contribution. A similar conclusion arises 
from measurements of D-meson production in Pb--Pb collisions~\cite{Adam:2015sza}.

\section{Conclusions} \label{section6}

The production of electrons from semileptonic decays of heavy-flavour hadrons has been measured 
at mid-rapidity ($|y| < 0.8$) in the $p_{\rm T}$ interval 0.5-3~GeV/$c$ in pp collisions 
and in 0--10\% and 20--40\% central Pb--Pb collisions at a centre-of-mass energy of 2.76 TeV per nucleon pair. 
The dominant background from photonic electron sources has been measured and subtracted via the 
photonic-electron tagging technique for the first time in pp and Pb--Pb collisions at the same 
energy. The systematic uncertainties have been substantially reduced (up to a factor 3), and the $p_{\rm T}$ coverage 
has been extended to lower values with respect to previously published ALICE measurements.

The measured nuclear modification factor $R_{\rm AA}$ of electrons from semileptonic heavy-flavour  
hadron decays confirms the strong suppression of high-$p_{\rm T}$ heavy-flavour hadrons in 
central Pb--Pb collisions with respect to the binary-collision scaled pp reference, consistent
with previous observations in various heavy-flavour channels.  
With decreasing $p_{\rm T}$, $R_{\rm AA}$ grows approaching values close to unity, 
as expected from the hypothesis of the binary-collision scaling for the total
heavy-quark yield. However, this kinematic region is sensitive to the effects of
nuclear shadowing:  the depletion of parton densities in nuclei
at low Bjorken $x$ values can reduce the heavy-quark production cross section per binary collision in Pb--Pb
with respect to the pp case. This initial-state effect is studied in p--Pb 
collisions~\cite{Adam:2015qda}. However, the present uncertainties on the $R_{\rm pPb}$ 
measurement do not allow quantitative conclusions on the modification of the PDF in nuclei in the low $p_{\rm T}$
region. With the improved precision of the results presented here, the Pb--Pb data 
exhibit their sensitivity to the modification of the PDF in nuclei, like nuclear shadowing, at low $p_{\rm T}$. The measured $R_{\rm AA}$ is in 
better agreement with TAMU and POWLANG model calculations when the nuclear modification of the PDF is included.

%
%

\newenvironment{acknowledgement}{\relax}{\relax}
\begin{acknowledgement}
\section*{Acknowledgements}

The ALICE Collaboration would like to thank all its engineers and technicians for their invaluable contributions to the construction of the experiment and the CERN accelerator teams for the outstanding performance of the LHC complex.
The ALICE Collaboration gratefully acknowledges the resources and support provided by all Grid centres and the Worldwide LHC Computing Grid (WLCG) collaboration.
The ALICE Collaboration acknowledges the following funding agencies for their support in building and running the ALICE detector:
A. I. Alikhanyan National Science Laboratory (Yerevan Physics Institute) Foundation (ANSL), State Committee of Science and World Federation of Scientists (WFS), Armenia;
Austrian Academy of Sciences and Nationalstiftung f\"{u}r Forschung, Technologie und Entwicklung, Austria;
Ministry of Communications and High Technologies, National Nuclear Research Center, Azerbaijan;
Conselho Nacional de Desenvolvimento Cient\'{\i}fico e Tecnol\'{o}gico (CNPq), Universidade Federal do Rio Grande do Sul (UFRGS), Financiadora de Estudos e Projetos (Finep) and Funda\c{c}\~{a}o de Amparo \`{a} Pesquisa do Estado de S\~{a}o Paulo (FAPESP), Brazil;
Ministry of Science \& Technology of China (MSTC), National Natural Science Foundation of China (NSFC) and Ministry of Education of China (MOEC) , China;
Ministry of Science and Education, Croatia;
Ministry of Education, Youth and Sports of the Czech Republic, Czech Republic;
The Danish Council for Independent Research | Natural Sciences, the Carlsberg Foundation and Danish National Research Foundation (DNRF), Denmark;
Helsinki Institute of Physics (HIP), Finland;
Commissariat \`{a} l'Energie Atomique (CEA) and Institut National de Physique Nucl\'{e}aire et de Physique des Particules (IN2P3) and Centre National de la Recherche Scientifique (CNRS), France;
Bundesministerium f\"{u}r Bildung, Wissenschaft, Forschung und Technologie (BMBF) and GSI Helmholtzzentrum f\"{u}r Schwerionenforschung GmbH, Germany;
General Secretariat for Research and Technology, Ministry of Education, Research and Religions, Greece;
National Research, Development and Innovation Office, Hungary;
Department of Atomic Energy Government of India (DAE), Department of Science and Technology, Government of India (DST), University Grants Commission, Government of India (UGC) and Council of Scientific and Industrial Research (CSIR), India;
Indonesian Institute of Science, Indonesia;
Centro Fermi - Museo Storico della Fisica e Centro Studi e Ricerche Enrico Fermi and Istituto Nazionale di Fisica Nucleare (INFN), Italy;
Institute for Innovative Science and Technology , Nagasaki Institute of Applied Science (IIST), Japan Society for the Promotion of Science (JSPS) KAKENHI and Japanese Ministry of Education, Culture, Sports, Science and Technology (MEXT), Japan;
Consejo Nacional de Ciencia (CONACYT) y Tecnolog\'{i}a, through Fondo de Cooperaci\'{o}n Internacional en Ciencia y Tecnolog\'{i}a (FONCICYT) and Direcci\'{o}n General de Asuntos del Personal Academico (DGAPA), Mexico;
Nederlandse Organisatie voor Wetenschappelijk Onderzoek (NWO), Netherlands;
The Research Council of Norway, Norway;
Commission on Science and Technology for Sustainable Development in the South (COMSATS), Pakistan;
Pontificia Universidad Cat\'{o}lica del Per\'{u}, Peru;
Ministry of Science and Higher Education and National Science Centre, Poland;
Korea Institute of Science and Technology Information and National Research Foundation of Korea (NRF), Republic of Korea;
Ministry of Education and Scientific Research, Institute of Atomic Physics and Romanian National Agency for Science, Technology and Innovation, Romania;
Joint Institute for Nuclear Research (JINR), Ministry of Education and Science of the Russian Federation and National Research Centre Kurchatov Institute, Russia;
Ministry of Education, Science, Research and Sport of the Slovak Republic, Slovakia;
National Research Foundation of South Africa, South Africa;
Centro de Aplicaciones Tecnol\'{o}gicas y Desarrollo Nuclear (CEADEN), Cubaenerg\'{\i}a, Cuba and Centro de Investigaciones Energ\'{e}ticas, Medioambientales y Tecnol\'{o}gicas (CIEMAT), Spain;
Swedish Research Council (VR) and Knut \& Alice Wallenberg Foundation (KAW), Sweden;
European Organization for Nuclear Research, Switzerland;
National Science and Technology Development Agency (NSDTA), Suranaree University of Technology (SUT) and Office of the Higher Education Commission under NRU project of Thailand, Thailand;
Turkish Atomic Energy Agency (TAEK), Turkey;
National Academy of  Sciences of Ukraine, Ukraine;
Science and Technology Facilities Council (STFC), United Kingdom;
National Science Foundation of the United States of America (NSF) and United States Department of Energy, Office of Nuclear Physics (DOE NP), United States of America.
\end{acknowledgement}

\bibliographystyle{utphys}
\bibliography{biblio}

\newpage
\appendix
\section{The ALICE Collaboration}
\label{app:collab}

\begingroup
\small
\begin{flushleft}
S.~Acharya\Irefn{org139}\And 
F.T.-.~Acosta\Irefn{org20}\And 
D.~Adamov\'{a}\Irefn{org93}\And 
J.~Adolfsson\Irefn{org80}\And 
M.M.~Aggarwal\Irefn{org98}\And 
G.~Aglieri Rinella\Irefn{org34}\And 
M.~Agnello\Irefn{org31}\And 
N.~Agrawal\Irefn{org48}\And 
Z.~Ahammed\Irefn{org139}\And 
S.U.~Ahn\Irefn{org76}\And 
S.~Aiola\Irefn{org144}
% ...
\And
\renewcommand\labelenumi{\textsuperscript{\theenumi}~}

\section*{Affiliation notes}
\renewcommand\theenumi{\roman{enumi}}
\begin{Authlist}
\item \Adef{org*}Deceased
\item \Adef{orgI}Dipartimento DET del Politecnico di Torino, Turin, Italy
\item \Adef{orgII}M.V. Lomonosov Moscow State University, D.V. Skobeltsyn Institute of Nuclear, Physics, Moscow, Russia
\item \Adef{orgIII}Department of Applied Physics, Aligarh Muslim University, Aligarh, India
\item \Adef{orgIV}Institute of Theoretical Physics, University of Wroclaw, Poland
\end{Authlist}

\section*{Collaboration Institutes}
\renewcommand\theenumi{\arabic{enumi}~}
\begin{Authlist}
\item \Idef{org1}A.I. Alikhanyan National Science Laboratory (Yerevan Physics Institute) Foundation, Yerevan, Armenia
\item \Idef{org2}Bogolyubov Institute for Theoretical Physics, National Academy of Sciences of Ukraine, Kiev, Ukraine
\item \Idef{org3}Bose Institute, Department of Physics  and Centre for Astroparticle Physics and Space Science (CAPSS), Kolkata, India
\item \Idef{org4}Budker Institute for Nuclear Physics, Novosibirsk, Russia
\item \Idef{org5}California Polytechnic State University, San Luis Obispo, California, United States
\item \Idef{org6}Central China Normal University, Wuhan, China
\item \Idef{org7}Centre de Calcul de l'IN2P3, Villeurbanne, Lyon, France
\item \Idef{org8}Centro de Aplicaciones Tecnol\'{o}gicas y Desarrollo Nuclear (CEADEN), Havana, Cuba
\item \Idef{org9}Centro de Investigaci\'{o}n y de Estudios Avanzados (CINVESTAV), Mexico City and M\'{e}rida, Mexico
\item \Idef{org10}Centro Fermi - Museo Storico della Fisica e Centro Studi e Ricerche ``Enrico Fermi', Rome, Italy
\item \Idef{org11}Chicago State University, Chicago, Illinois, United States
\item \Idef{org12}China Institute of Atomic Energy, Beijing, China
\item \Idef{org13}Chonbuk National University, Jeonju, Republic of Korea
\item \Idef{org14}Comenius University Bratislava, Faculty of Mathematics, Physics and Informatics, Bratislava, Slovakia
\item \Idef{org15}COMSATS Institute of Information Technology (CIIT), Islamabad, Pakistan
\item \Idef{org16}Creighton University, Omaha, Nebraska, United States
\item \Idef{org17}Department of Physics, Aligarh Muslim University, Aligarh, India
\item \Idef{org18}Department of Physics, Pusan National University, Pusan, Republic of Korea
\item \Idef{org19}Department of Physics, Sejong University, Seoul, Republic of Korea
\item \Idef{org20}Department of Physics, University of California, Berkeley, California, United States
\item \Idef{org21}Department of Physics, University of Oslo, Oslo, Norway
\item \Idef{org22}Department of Physics and Technology, University of Bergen, Bergen, Norway
\item \Idef{org23}Dipartimento di Fisica dell'Universit\`{a} 'La Sapienza' and Sezione INFN, Rome, Italy
\item \Idef{org24}Dipartimento di Fisica dell'Universit\`{a} and Sezione INFN, Cagliari, Italy
\item \Idef{org25}Dipartimento di Fisica dell'Universit\`{a} and Sezione INFN, Trieste, Italy
\item \Idef{org26}Dipartimento di Fisica dell'Universit\`{a} and Sezione INFN, Turin, Italy
\item \Idef{org27}Dipartimento di Fisica e Astronomia dell'Universit\`{a} and Sezione INFN, Bologna, Italy
\item \Idef{org28}Dipartimento di Fisica e Astronomia dell'Universit\`{a} and Sezione INFN, Catania, Italy
\item \Idef{org29}Dipartimento di Fisica e Astronomia dell'Universit\`{a} and Sezione INFN, Padova, Italy
\item \Idef{org30}Dipartimento di Fisica `E.R.~Caianiello' dell'Universit\`{a} and Gruppo Collegato INFN, Salerno, Italy
\item \Idef{org31}Dipartimento DISAT del Politecnico and Sezione INFN, Turin, Italy
\item \Idef{org32}Dipartimento di Scienze e Innovazione Tecnologica dell'Universit\`{a} del Piemonte Orientale and INFN Sezione di Torino, Alessandria, Italy
\item \Idef{org33}Dipartimento Interateneo di Fisica `M.~Merlin' and Sezione INFN, Bari, Italy
\item \Idef{org34}European Organization for Nuclear Research (CERN), Geneva, Switzerland
\item \Idef{org35}Faculty of Electrical Engineering, Mechanical Engineering and Naval Architecture, University of Split, Split, Croatia
\item \Idef{org36}Faculty of Engineering and Science, Western Norway University of Applied Sciences, Bergen, Norway
\item \Idef{org37}Faculty of Nuclear Sciences and Physical Engineering, Czech Technical University in Prague, Prague, Czech Republic
\item \Idef{org38}Faculty of Science, P.J.~\v{S}af\'{a}rik University, Ko\v{s}ice, Slovakia
\item \Idef{org39}Frankfurt Institute for Advanced Studies, Johann Wolfgang Goethe-Universit\"{a}t Frankfurt, Frankfurt, Germany
\item \Idef{org40}Gangneung-Wonju National University, Gangneung, Republic of Korea
\item \Idef{org41}Gauhati University, Department of Physics, Guwahati, India
\item \Idef{org42}Helmholtz-Institut f\"{u}r Strahlen- und Kernphysik, Rheinische Friedrich-Wilhelms-Universit\"{a}t Bonn, Bonn, Germany
\item \Idef{org43}Helsinki Institute of Physics (HIP), Helsinki, Finland
\item \Idef{org44}High Energy Physics Group,  Universidad Aut\'{o}noma de Puebla, Puebla, Mexico
\item \Idef{org45}Hiroshima University, Hiroshima, Japan
\item \Idef{org46}Hochschule Worms, Zentrum  f\"{u}r Technologietransfer und Telekommunikation (ZTT), Worms, Germany
\item \Idef{org47}Horia Hulubei National Institute of Physics and Nuclear Engineering, Bucharest, Romania
\item \Idef{org48}Indian Institute of Technology Bombay (IIT), Mumbai, India
\item \Idef{org49}Indian Institute of Technology Indore, Indore, India
\item \Idef{org50}Indonesian Institute of Sciences, Jakarta, Indonesia
\item \Idef{org51}INFN, Laboratori Nazionali di Frascati, Frascati, Italy
\item \Idef{org52}INFN, Sezione di Bari, Bari, Italy
\item \Idef{org53}INFN, Sezione di Bologna, Bologna, Italy
\item \Idef{org54}INFN, Sezione di Cagliari, Cagliari, Italy
\item \Idef{org55}INFN, Sezione di Catania, Catania, Italy
\item \Idef{org56}INFN, Sezione di Padova, Padova, Italy
\item \Idef{org57}INFN, Sezione di Roma, Rome, Italy
\item \Idef{org58}INFN, Sezione di Torino, Turin, Italy
\item \Idef{org59}INFN, Sezione di Trieste, Trieste, Italy
\item \Idef{org60}Inha University, Incheon, Republic of Korea
\item \Idef{org61}Institut de Physique Nucl\'{e}aire d'Orsay (IPNO), Institut National de Physique Nucl\'{e}aire et de Physique des Particules (IN2P3/CNRS), Universit\'{e} de Paris-Sud, Universit\'{e} Paris-Saclay, Orsay, France
\item \Idef{org62}Institute for Nuclear Research, Academy of Sciences, Moscow, Russia
\item \Idef{org63}Institute for Subatomic Physics, Utrecht University/Nikhef, Utrecht, Netherlands
\item \Idef{org64}Institute for Theoretical and Experimental Physics, Moscow, Russia
\item \Idef{org65}Institute of Experimental Physics, Slovak Academy of Sciences, Ko\v{s}ice, Slovakia
\item \Idef{org66}Institute of Physics, Homi Bhabha National Institute, Bhubaneswar, India
\item \Idef{org67}Institute of Physics of the Czech Academy of Sciences, Prague, Czech Republic
\item \Idef{org68}Institute of Space Science (ISS), Bucharest, Romania
\item \Idef{org69}Institut f\"{u}r Kernphysik, Johann Wolfgang Goethe-Universit\"{a}t Frankfurt, Frankfurt, Germany
\item \Idef{org70}Instituto de Ciencias Nucleares, Universidad Nacional Aut\'{o}noma de M\'{e}xico, Mexico City, Mexico
\item \Idef{org71}Instituto de F\'{i}sica, Universidade Federal do Rio Grande do Sul (UFRGS), Porto Alegre, Brazil
\item \Idef{org72}Instituto de F\'{\i}sica, Universidad Nacional Aut\'{o}noma de M\'{e}xico, Mexico City, Mexico
\item \Idef{org73}iThemba LABS, National Research Foundation, Somerset West, South Africa
\item \Idef{org74}Johann-Wolfgang-Goethe Universit\"{a}t Frankfurt Institut f\"{u}r Informatik, Fachbereich Informatik und Mathematik, Frankfurt, Germany
\item \Idef{org75}Joint Institute for Nuclear Research (JINR), Dubna, Russia
\item \Idef{org76}Korea Institute of Science and Technology Information, Daejeon, Republic of Korea
\item \Idef{org77}KTO Karatay University, Konya, Turkey
\item \Idef{org78}Laboratoire de Physique Subatomique et de Cosmologie, Universit\'{e} Grenoble-Alpes, CNRS-IN2P3, Grenoble, France
\item \Idef{org79}Lawrence Berkeley National Laboratory, Berkeley, California, United States
\item \Idef{org80}Lund University Department of Physics, Division of Particle Physics, Lund, Sweden
\item \Idef{org81}Nagasaki Institute of Applied Science, Nagasaki, Japan
\item \Idef{org82}Nara Women{'}s University (NWU), Nara, Japan
\item \Idef{org83}National and Kapodistrian University of Athens, School of Science, Department of Physics , Athens, Greece
\item \Idef{org84}National Centre for Nuclear Research, Warsaw, Poland
\item \Idef{org85}National Institute of Science Education and Research, Homi Bhabha National Institute, Jatni, India
\item \Idef{org86}National Nuclear Research Center, Baku, Azerbaijan
\item \Idef{org87}National Research Centre Kurchatov Institute, Moscow, Russia
\item \Idef{org88}Niels Bohr Institute, University of Copenhagen, Copenhagen, Denmark
\item \Idef{org89}Nikhef, National institute for subatomic physics, Amsterdam, Netherlands
\item \Idef{org90}NRC Kurchatov Institute IHEP, Protvino, Russia
\item \Idef{org91}NRNU Moscow Engineering Physics Institute, Moscow, Russia
\item \Idef{org92}Nuclear Physics Group, STFC Daresbury Laboratory, Daresbury, United Kingdom
\item \Idef{org93}Nuclear Physics Institute of the Czech Academy of Sciences, \v{R}e\v{z} u Prahy, Czech Republic
\item \Idef{org94}Oak Ridge National Laboratory, Oak Ridge, Tennessee, United States
\item \Idef{org95}Ohio State University, Columbus, Ohio, United States
\item \Idef{org96}Petersburg Nuclear Physics Institute, Gatchina, Russia
\item \Idef{org97}Physics department, Faculty of science, University of Zagreb, Zagreb, Croatia
\item \Idef{org98}Physics Department, Panjab University, Chandigarh, India
\item \Idef{org99}Physics Department, University of Jammu, Jammu, India
\item \Idef{org100}Physics Department, University of Rajasthan, Jaipur, India
\item \Idef{org101}Physikalisches Institut, Eberhard-Karls-Universit\"{a}t T\"{u}bingen, T\"{u}bingen, Germany
\item \Idef{org102}Physikalisches Institut, Ruprecht-Karls-Universit\"{a}t Heidelberg, Heidelberg, Germany
\item \Idef{org103}Physik Department, Technische Universit\"{a}t M\"{u}nchen, Munich, Germany
\item \Idef{org104}Research Division and ExtreMe Matter Institute EMMI, GSI Helmholtzzentrum f\"ur Schwerionenforschung GmbH, Darmstadt, Germany
\item \Idef{org105}Rudjer Bo\v{s}kovi\'{c} Institute, Zagreb, Croatia
\item \Idef{org106}Russian Federal Nuclear Center (VNIIEF), Sarov, Russia
\item \Idef{org107}Saha Institute of Nuclear Physics, Homi Bhabha National Institute, Kolkata, India
\item \Idef{org108}School of Physics and Astronomy, University of Birmingham, Birmingham, United Kingdom
\item \Idef{org109}Secci\'{o}n F\'{\i}sica, Departamento de Ciencias, Pontificia Universidad Cat\'{o}lica del Per\'{u}, Lima, Peru
\item \Idef{org110}Shanghai Institute of Applied Physics, Shanghai, China
\item \Idef{org111}St. Petersburg State University, St. Petersburg, Russia
\item \Idef{org112}Stefan Meyer Institut f\"{u}r Subatomare Physik (SMI), Vienna, Austria
\item \Idef{org113}SUBATECH, IMT Atlantique, Universit\'{e} de Nantes, CNRS-IN2P3, Nantes, France
\item \Idef{org114}Suranaree University of Technology, Nakhon Ratchasima, Thailand
\item \Idef{org115}Technical University of Ko\v{s}ice, Ko\v{s}ice, Slovakia
\item \Idef{org116}Technische Universit\"{a}t M\"{u}nchen, Excellence Cluster 'Universe', Munich, Germany
\item \Idef{org117}The Henryk Niewodniczanski Institute of Nuclear Physics, Polish Academy of Sciences, Cracow, Poland
\item \Idef{org118}The University of Texas at Austin, Austin, Texas, United States
\item \Idef{org119}Universidad Aut\'{o}noma de Sinaloa, Culiac\'{a}n, Mexico
\item \Idef{org120}Universidade de S\~{a}o Paulo (USP), S\~{a}o Paulo, Brazil
\item \Idef{org121}Universidade Estadual de Campinas (UNICAMP), Campinas, Brazil
\item \Idef{org122}Universidade Federal do ABC, Santo Andre, Brazil
\item \Idef{org123}University College of Southeast Norway, Tonsberg, Norway
\item \Idef{org124}University of Cape Town, Cape Town, South Africa
\item \Idef{org125}University of Houston, Houston, Texas, United States
\item \Idef{org126}University of Jyv\"{a}skyl\"{a}, Jyv\"{a}skyl\"{a}, Finland
\item \Idef{org127}University of Liverpool, Liverpool, United Kingdom
\item \Idef{org128}University of Tennessee, Knoxville, Tennessee, United States
\item \Idef{org129}University of the Witwatersrand, Johannesburg, South Africa
\item \Idef{org130}University of Tokyo, Tokyo, Japan
\item \Idef{org131}University of Tsukuba, Tsukuba, Japan
\item \Idef{org132}Universit\'{e} Clermont Auvergne, CNRS/IN2P3, LPC, Clermont-Ferrand, France
\item \Idef{org133}Universit\'{e} de Lyon, Universit\'{e} Lyon 1, CNRS/IN2P3, IPN-Lyon, Villeurbanne, Lyon, France
\item \Idef{org134}Universit\'{e} de Strasbourg, CNRS, IPHC UMR 7178, F-67000 Strasbourg, France, Strasbourg, France
\item \Idef{org135} Universit\'{e} Paris-Saclay Centre d¿\'Etudes de Saclay (CEA), IRFU, Department de Physique Nucl\'{e}aire (DPhN), Saclay, France
\item \Idef{org136}Universit\`{a} degli Studi di Foggia, Foggia, Italy
\item \Idef{org137}Universit\`{a} degli Studi di Pavia, Pavia, Italy
\item \Idef{org138}Universit\`{a} di Brescia, Brescia, Italy
\item \Idef{org139}Variable Energy Cyclotron Centre, Homi Bhabha National Institute, Kolkata, India
\item \Idef{org140}Warsaw University of Technology, Warsaw, Poland
\item \Idef{org141}Wayne State University, Detroit, Michigan, United States
\item \Idef{org142}Westf\"{a}lische Wilhelms-Universit\"{a}t M\"{u}nster, Institut f\"{u}r Kernphysik, M\"{u}nster, Germany
\item \Idef{org143}Wigner Research Centre for Physics, Hungarian Academy of Sciences, Budapest, Hungary
\item \Idef{org144}Yale University, New Haven, Connecticut, United States
\item \Idef{org145}Yonsei University, Seoul, Republic of Korea
\end{Authlist}
\endgroup
\end{document}